%% file: main.tex
\documentclass[aps,prc,reprint,floatfix,superscriptaddress]{revtex4-2}
\usepackage{xcolor}
\usepackage{graphicx,amsfonts,amsmath,amssymb}
\usepackage{booktabs,threeparttable,siunitx,multirow}
\usepackage[colorlinks=true,linkcolor=blue,citecolor=blue,urlcolor=blue]{hyperref}
\usepackage{orcidlink}

\begin{document}

\title{Physics-structured cooperative neural network for baseline-free nuclear mass modeling}

\author{Peiwen Zai\,\orcidlink{0009-0004-6269-7090}}
\email[]{pewenz@mail.bnu.edu.cn}
\affiliation{The Key Laboratory of Beam Technology of Ministry of Education, School of Physics and Astronomy, Beijing Normal University, Beijing 100875, China}

\author{Wei Cheng}
\affiliation{The Key Laboratory of Beam Technology of Ministry of Education, School of Physics and Astronomy, Beijing Normal University, Beijing 100875, China}
\affiliation{Advanced Institute of Natural Sciences, Beijing Normal University at Zhuhai, Zhuhai 519087, China}

\author{Feng-Shou Zhang\,\orcidlink{0000-0003-0507-0983}}
\email[]{fszhang@bnu.edu.cn}
\affiliation{The Key Laboratory of Beam Technology of Ministry of Education, School of Physics and Astronomy, Beijing Normal University, Beijing 100875, China}
\affiliation{Institute of Radiation Technology, Beijing Academy of Science and Technology, Beijing 100875, China}
\affiliation{Center of Theoretical Nuclear Physics, National Laboratory of Heavy Ion Accelerator of Lanzhou, Lanzhou 730000, China}

\date{\today}

\begin{abstract}
Machine learning approaches can improve nuclear mass modelling, but the most accurate strategies often depend on a theoretical mass baseline or hand-crafted physics features.
We test whether a modular architecture encoding selected nuclear-structure priors improves baseline-free direct prediction and yields informative branch diagnostics.
The Cooperative Neural Network (CoNN) implements this approach through four form-constrained branches: a smooth macroscopic network, discrete embeddings, a two-dimensional regional grid, and a parity-aware network.
It extracts complementary patterns from $(Z,N)$ through these branches and sums their outputs to predict binding energies without a theoretical mass-model baseline.
Thus, the model retains physics priors while reducing its reliance on engineered input features.
On AME2020, CoNN reaches a root-mean-square deviation (RMSD) of 0.269\,MeV for 3558 nuclei, compared with 0.836\,MeV for a parameter-matched unstructured MLP.
It also gives RMSDs of 0.419\,MeV on a held-out interpolation subset and 0.728\,MeV on 122 nuclei newly measured since AME2016.
The learned branch outputs show recognizable physical patterns, including embedding shell-kink signatures at major magic numbers and odd--even staggering along isotopic chains.
These results identify architecture-level priors as a practical route to baseline-free mass prediction, with learned components that help diagnose both nuclear-structure patterns and extrapolation limits.
\end{abstract}

\maketitle

\input{content.tex}

\begin{acknowledgments}
This work was supported by the National Natural Science Foundation of China under Grants No. 12135004.
\end{acknowledgments}

\bibliography{ref.bib}

\end{document}

%% file: content.tex
\section{Introduction}\label{sec:introduction}

Nuclear binding energies determine nuclear stability, decay energetics, and reaction thresholds\,\cite{2003Lunney+Recent,2016Mumpower+Impact}.
Reliable mass predictions are therefore essential far from stability, especially for modeling the path of the rapid neutron-capture process ($r$-process) through neutron-rich nuclei beyond current experimental reach\,\cite{2016Martin+Impact,2021Cowan+Origin}.
The latest Atomic Mass Evaluation (AME2020)\,\cite{2021Wang+AME} reports binding energies for 3558 nuclides.
Theoretical estimates range from about 7000\,\cite{2020Neufcourt+Quantified,2012Erler+Limits,2004Thoennessen+Reaching} to around 9000\,\cite{2018Xia+Limits} bound nuclei, leaving much of the nuclear chart to theoretical or data-driven extrapolation.

Traditional global mass models can be broadly grouped into microscopic and macroscopic--microscopic approaches\,\cite{2003Lunney+Recent, 2018Sobiczewski+Detailed}.
Microscopic models based on nuclear density functional theory (DFT)\,\cite{2003Bender+Selfconsistent, 2006Meng+Relativistic,2009Goriely+First,2016Goriely+Further,2025Grams+Skyrme,2026Grams+SkyrmeHartreeFockBogoliubov}, including Skyrme/Gogny Hartree--Fock--Bogoliubov and covariant mean-field formulations, provide a self-consistent description of nuclear structure across the chart.
Their predictive accuracy is typically limited to root-mean-square deviations (RMSDs) of 0.5--0.8\,MeV\,\cite{2018Sobiczewski+Detailed}, reflecting both the incomplete treatment of many-body correlations and the constraints of current energy-density functionals.
Macroscopic--microscopic models instead construct a deformation-dependent energy from a smooth macroscopic term and microscopic corrections based on a phenomenological single-particle spectrum\,\cite{1967Strutinsky+Shell,2016Moller+Nuclear,2014Wang+Surface,1972BRACK+Funny}.
Ground-state masses follow from the minimum of this energy surface, while the treatments of pairing, residual correlations, and deformation vary among implementations.
Representative implementations such as FRDM2012\,\cite{2016Moller+Nuclear} and WS4\,\cite{2014Wang+Surface} achieve RMSDs of 0.3--0.6\,MeV.
This construction motivates the organizing principle used here: representing the mass surface as a smooth bulk trend plus structured microscopic corrections.

In recent years, machine-learning (ML) methods have become effective complements to traditional mass models\,\cite{2019Carleo+Machine, 2022Boehnlein+Colloquium}.
One successful strategy is residual correction, in which an ML model is trained on the difference $\Delta B = B_{\mathrm{exp}} - B_{\mathrm{th}}$ between experiment and a chosen theoretical baseline.
The model therefore learns systematic deficiencies of that baseline rather than the full binding-energy surface.
Using Bayesian neural networks\,\cite{2016Utama+Nuclear,2017Utama+Refining, 2018Neufcourt+Bayesian,2018Niu+Nuclear,2022Niu+Nuclear}, kernel methods\,\cite{2022Wu+Multitask, 2024Yuksel+Nuclear}, and various deep architectures\,\cite{2025Lu+Nuclear,2025Huang+Validation,2025Jalili+Deep}, such hybrid approaches have achieved RMSDs of 80--170\,keV.
However, this accuracy is tied to the external mass model, so these methods are baseline-dependent corrections rather than standalone predictive frameworks.

Direct prediction from the minimal identifiers $(Z,N)$, without a theoretical baseline, poses a different problem.
The model must represent the full binding-energy scale, spanning ${\sim}2000$\,MeV\,\cite{2021Wang+AME}, rather than only residuals around a physics model.
Standard feed-forward networks in this setting have exhibited RMSDs at the MeV level, even when the number of parameters is substantially increased\,\cite{2022Lovell+Nuclear,2004Athanassopoulos+Nuclear}.
This difficulty reflects the multi-scale structure of the mass surface, which a single unstructured architecture must learn simultaneously\,\cite{2026Huang+Principal}.
Subsequent work has reduced RMSDs to 0.2--0.5\,MeV by augmenting $(Z,N)$ with physics-motivated features, including parity indicators, distances to shell closures, and isospin-asymmetry ratios\,\cite{2022Mumpower+Physically, 2024Zeng+Nuclear}.
This route is effective, but the chosen descriptors determine which nuclear structures are made explicit in the input representation\,\cite{2022Lovell+Nuclear,2024Zeng+Nuclear,2025Huang+Validation}.

This motivates a different member of the broader family of physics-structured, architecture-constrained learning models, in which domain organization is built into modules, decompositions, shared tasks, or constrained functional forms.
Related examples include data-driven density-functional models for nuclei, multi-task learning of masses and separation energies, and decomposed neural-network potentials in atomistic systems\,\cite{2007Behler+Generalized,2017Smith+ANI1,2022Wu+Multitask,2024Yang+Datadriven}.
Within this established category, we examine whether selected nuclear-structure priors can be moved from external input descriptors into the architecture of a baseline-free direct predictor.
Guided by the macroscopic--microscopic picture, we introduce the Cooperative Neural Network (CoNN), which predicts binding energies from $(Z,N)$ and writes the output as a sum of four constrained branches: a smooth macroscopic network, discrete embeddings, a two-dimensional regional grid, and a parity-aware network.
Thus, the model uses explicit architectural priors, including parity extraction and finite embedding and grid domains, but does not use a theoretical mass-model baseline or external descriptors such as shell distances and isospin-asymmetry ratios.

The evaluation addresses whether architectural constraints improve baseline-free direct prediction beyond model size and which nuclear-mass patterns appear in the correction-branch outputs.
A parameter-matched MLP provides a controlled baseline, while published direct-prediction models provide broader performance context.
Branch ablations and output diagnostics examine odd--even staggering, embedding shell-kink signatures, and nonseparable regional corrections, while controlled model variants test shell-kink stability.
Derived quantities and neutron-drip-line predictions then probe local mass-surface structure, extrapolation behavior, and architectural failure modes, with physics-based mass models serving as references.

The paper is organized as follows.
Section~\ref{sec:methodology} describes the dataset, modular architecture, and alternating training procedure.
Section~\ref{sec:results} benchmarks interpolation and extrapolation performance against representative ML approaches, analyzes the learned branches, and uses physics-based mass models as references for derived-quantity diagnostics.
Section~\ref{sec:conclusion} summarizes the scope, limitations, and possible extensions of the approach.

\section{Methodology}\label{sec:methodology}

CoNN is an additive, architecture-constrained predictor of nuclear binding energies from $(Z,N)$ directly.
It writes the model output as the sum of a smooth macroscopic network and three form-constrained correction branches: discrete embeddings, a regional grid, and a parity-aware network.
Together, they encode continuous, index-wise, joint two-dimensional, and parity-dependent functions of $(Z,N)$.
The corresponding forms are motivated by bulk trends, shell structure, regional correlations, and odd--even staggering in nuclear masses.
In the following, we describe the dataset (Sec.~\ref{sec:data}), the branch definitions and their physical motivation (Sec.~\ref{sec:architecture}), and the alternating training procedure used to encourage a division of labor between macroscopic and microscopic terms (Sec.~\ref{sec:training}).

\subsection{Data}\label{sec:data}

\begin{figure}[!htb]
\centering
\includegraphics[width=\columnwidth]{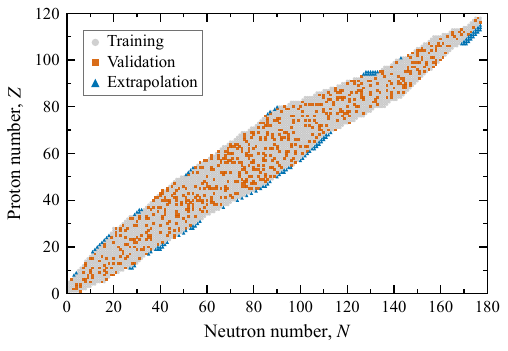}
\caption{Distribution of the training ($n=2748$, gray circles), validation ($n=688$, orange squares), and extrapolation ($n=122$, blue triangles) subsets on the $(N,Z)$ chart. The training and validation subsets form an 80:20 random split of the AME2016-overlap set; the extrapolation subset contains AME2020 nuclei absent from AME2016.}
\label{fig:distribution}
\end{figure}

We use binding energies from AME2020\,\cite{2021Wang+AME} as reference labels.
We construct the split in two steps.
First, measurement history separates the 3558 AME2020 nuclei into an AME2016-overlap set of 3436 nuclei\,\cite{2017Wang+AME2016} and an AME2020-new set of 122 nuclei.
Second, only the AME2016-overlap set is divided randomly into training and validation subsets in an 80:20 ratio.
The validation subset tests interpolation within the previously known region.
The AME2020-new set is held out as a temporal out-of-sample test of near-boundary extrapolation to newly measured nuclei.
Figure~\ref{fig:distribution} shows the three subsets on the $(N,Z)$ chart.

\subsection{Model architecture}\label{sec:architecture}

\begin{figure*}[!htb]
\centering
\includegraphics[width=\textwidth]{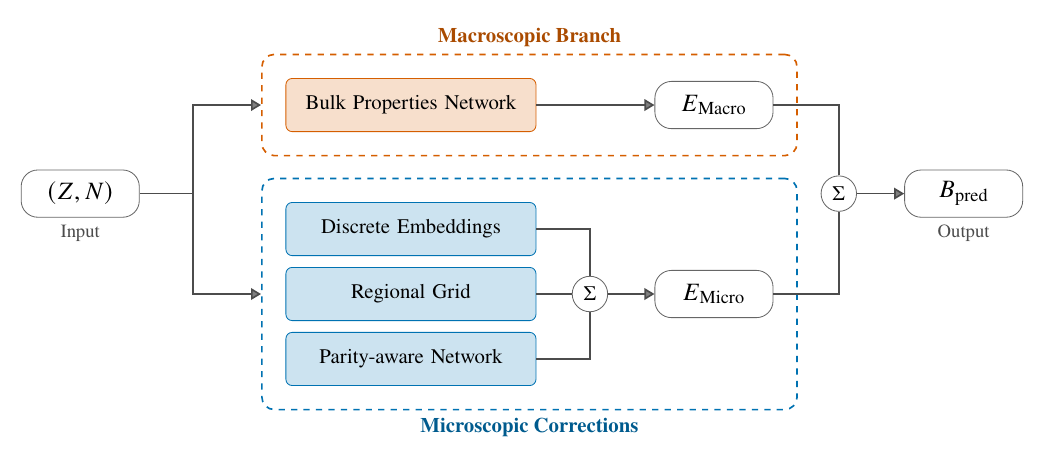}
\caption{Architecture of the CoNN model. The model output is written as the sum of a smooth macroscopic branch and three constrained correction branches: discrete embeddings, a regional grid, and a parity-aware network. The branch outputs are added to yield $B_{\mathrm{pred}}$.}
\label{fig:architecture}
\end{figure*}

Following the macroscopic--microscopic picture, CoNN represents the predicted binding energy as an additive model output:
\begin{equation}
\begin{aligned}
B_{\mathrm{pred}} &= E_{\mathrm{Macro}} + E_{\mathrm{Micro}},\\
E_{\mathrm{Micro}} &= E_{\mathrm{emb}} + E_{\mathrm{grid}} + E_{\mathrm{pair}}.
\end{aligned}
\label{eq:decomposition}
\end{equation}
Here $E_{\mathrm{Macro}}$ is the output of a smooth bulk branch.
$E_{\mathrm{emb}}$, $E_{\mathrm{grid}}$, and $E_{\mathrm{pair}}$ are correction-branch outputs defined by discrete lookup, two-dimensional interpolation, and parity-aware functional forms, respectively.
These choices impose different forms of dependence on $(Z,N)$ for the four branch outputs.
Specifically, $E_{\mathrm{Macro}}$ is parameterized by a fully connected network that receives coordinates $(Z,N)$ as two min--max-scaled scalar inputs.
$E_{\mathrm{emb}}$ uses discrete scalar embeddings indexed by $Z$ and $N$; $E_{\mathrm{grid}}$ uses a learnable two-dimensional grid with bilinear interpolation; and $E_{\mathrm{pair}}$ uses a small parity-aware network.
The overall architecture is illustrated in Fig.~\ref{fig:architecture}.

\paragraph*{Smooth macroscopic network.}
The largest-scale variation of nuclear binding energies is smooth in $(Z,N)$, as reflected in liquid-drop descriptions\,\cite{1966Myers+Nuclear,2003Lunney+Recent}.
We model this component with a fully connected encoder--decoder network:
\begin{equation}
    E_{\mathrm{Macro}}(Z, N) = \mathcal{D}(\mathcal{E}(Z, N)),
\end{equation}
where $\mathcal{E}$ maps $(Z,N)$ to a 16-dimensional representation through three hidden layers of width 128 with LeakyReLU activations.
The decoder $\mathcal{D}$ uses the mirrored layer structure and maps this representation back to a scalar.
Relative to the width-128 hidden layers, the 16-dimensional central representation acts as a bottleneck that reduces the effective capacity of this branch\,\cite{2006Hinton+Reducing}.
Together with the spectral bias of gradient-based optimization, which favors low-frequency functions in fully connected networks\,\cite{2019Rahaman+Spectral, 2020Xu+Frequency}, this design gives $E_{\mathrm{Macro}}$ the role of the continuous, slowly varying branch in the additive model.
The discrete-embedding, regional-grid, and parity-aware branches then add discrete, local, and parity-dependent corrections to this smooth component.

\paragraph*{Discrete embeddings.}
Shell closures introduce localized changes in the mass surface near magic proton and neutron numbers that are difficult to represent with a purely smooth branch\,\cite{1967Strutinsky+Shell,2003Lunney+Recent,2024Buskirk+Nucleonic}. 
Motivated by this $Z$- and $N$-indexed structure, the embedding branch uses trainable scalar embeddings:
\begin{equation}
    E_{\mathrm{emb}}(Z, N) = e_Z[Z] + e_N[N],
    \label{eq:shell}
\end{equation}
Here $e_Z$ is a lookup table containing one scalar parameter for each integer proton number in the model domain.
$e_N$ is the analogous table for neutron number.
The bracket notation denotes lookup: $e_Z[Z]$ selects the scalar whose table index is $Z$, and $e_N[N]$ selects the scalar whose table index is $N$.
This additive, separable form is motivated by the independent-particle picture in which proton and neutron shell closures arise from separate single-particle spectra\,\cite{1949Mayer+Closed,1949Haxel+Magic}. 
Because this form is separable, it cannot represent residual structures that depend jointly on $Z$ and $N$.
The regional-grid branch below supplies this two-dimensional dependence.

\paragraph*{Regional grid.}
To include a branch with joint dependence on the two nucleon numbers, we introduce a learnable two-dimensional parameter grid $\mathbf{G} \in \mathbb{R}^{H \times W}$, with $H=50$ and $W=60$ grid nodes along the $Z$ and $N$ directions.
For a nucleus $(Z,N)$, the integer coordinates are first mapped to normalized grid coordinates
\begin{equation}
    z_g = 2Z/Z_{\max}-1,\qquad n_g = 2N/N_{\max}-1,
    \label{eq:grid_norm}
\end{equation}
with $z_g,n_g\in[-1,1]$ over the finite model domain.
The regional-grid output is then
\begin{equation}
    E_{\mathrm{grid}}(Z, N)
    = \mathrm{Interp}\!\left(\mathbf{G},\, z_g,\, n_g \right),
    \label{eq:cor}
\end{equation}
where $\mathrm{Interp}$ denotes bilinear interpolation on the grid.
Specifically, let
\begin{equation}
    u=\frac{z_g+1}{2}(H-1),\qquad
    v=\frac{n_g+1}{2}(W-1),
\end{equation}
and let $i=\lfloor u\rfloor$, $j=\lfloor v\rfloor$, $\alpha=u-i$, and $\beta=v-j$.
Then, away from the upper grid boundary,
\begin{equation}
\begin{aligned}
\mathrm{Interp}(\mathbf{G},z_g,n_g)
&=(1-\alpha)(1-\beta)G_{i,j}
  +\alpha(1-\beta)G_{i+1,j}\\
&\quad +(1-\alpha)\beta G_{i,j+1}
  +\alpha\beta G_{i+1,j+1},
\end{aligned}
\label{eq:bilinear_interp}
\end{equation}
with boundary indices clipped to the grid domain.
Because neighboring nuclei share interpolated grid values, the finite grid resolution gives this branch nonseparable regional variation while avoiding an independent parameter for each nucleus.
The finite model domain, the normalization by $Z_{\max}$ and $N_{\max}$, and the grid resolution $(H,W)$ define the coordinate system, range, and spatial scale of the learned regional correction.

\paragraph*{Parity-aware network.}
Nuclear pairing correlations produce odd--even staggering in binding energies, with a characteristic dependence on the parities of $Z$ and $N$ and a typical scale of order 1\,MeV\,\cite{1958Bohr+Possible}.
The preceding branches do not explicitly condition on nucleon-number parity.
We therefore model this parity-dependent correction with a small multilayer perceptron (MLP; one hidden layer of width 16 with SiLU activation):
\begin{equation}
    E_{\mathrm{pair}}(Z, N)
    = \mathrm{MLP}\!\left(\left[
      \pi_Z,\, \pi_N,\,
      \tfrac{Z}{Z_{\max}},\, \tfrac{N}{N_{\max}}
    \right]\right),
    \label{eq:pair}
\end{equation}
Here $\pi_Z = Z \bmod 2$ and $\pi_N = N \bmod 2$ are parity indicators for the four parity classes: even--even, even--odd, odd--even, and odd--odd.
The normalized coordinates $Z/Z_{\max}$ and $N/N_{\max}$ provide position dependence within the finite model domain, allowing the parity-dependent correction to vary across the chart.
The parity indicators are deterministic functions of $(Z,N)$.
This design provides a dedicated functional form for known modulo-two organization.

The complete model has approximately $7.4\times 10^4$ trainable parameters, with ${\sim}71{,}000$ in the smooth macroscopic network and ${\sim}3{,}400$ in the three correction branches.
To control for model size, we also train a plain MLP baseline with a matched parameter count (${\sim}7.4\times 10^4$; 8 hidden layers of width 102).
This baseline uses the same dataset and number of training epochs but has no branch structure or form constraints.

\begin{table*}[!t]
\centering
\begin{threeparttable}
\caption{Baseline-free direct-prediction ML approaches to nuclear mass modeling. Overall RMSDs are evaluated on the dataset listed in each row. Where available, extrapolation RMSDs are evaluated on the same 122-nucleus extrapolation set defined by the AME2016-to-AME2020 measurement-history split in Sec.~\ref{sec:data}. ``Physics prior'' denotes explicit nuclear-structure information, not generic learning-algorithm inductive biases. ``Features'' counts explicit input features, and ``Data'' gives the AME edition used for the overall RMSD.}
\label{tab:comparison}
\setlength{\tabcolsep}{8pt}
\renewcommand{\arraystretch}{1.15}
\begin{tabular}{@{} l c c @{\hspace{10pt}} S[table-format=1.3] c @{\hspace{10pt}} S[table-format=1.3] c c @{}}
\toprule
\multirow{2}{*}{\textbf{Model}} &
\multirow{2}{*}{\textbf{Physics Prior}} &
\multirow{2}{*}{\textbf{Features}} &
\multicolumn{2}{c@{\hspace{10pt}}}{\textbf{Overall Accuracy}} &
\multicolumn{2}{c}{\textbf{Extrapolation}} &
\multirow{2}{*}{\textbf{Data}} \\
\cmidrule(lr){4-5} \cmidrule(lr){6-7}
& & &
\multicolumn{1}{c}{$\boldsymbol{\sigma}_{\textbf{rms}}$ \textbf{(MeV)}} &
\textbf{Nuclei} &
\multicolumn{1}{c}{$\boldsymbol{\sigma}_{\textbf{rms}}$ \textbf{(MeV)}} &
\textbf{Nuclei} & \\
\midrule
ANN2\,\cite{2024Zeng+Nuclear}          & none       &  2 & 1.180 & 3556 & 1.050 & 122 & AME2016 \\
KAN-2\,\cite{2025Liu+Kolmogorovarnold} & none       &  2 & 0.870 & 3456 & \textemdash & \textemdash & AME2020 \\
MLP\tnote{a}                           & none       &  2 & 0.836 & 3558 & 1.232 & 122 & AME2020 \\
ANN7\,\cite{2024Zeng+Nuclear}          & input descriptors   &  7 & 0.200 & 3556 & 0.340 & 122 & AME2016 \\
KAN-11\,\cite{2025Liu+Kolmogorovarnold} & input descriptors   & 11 & 0.260 & 3456 & \textemdash & \textemdash & AME2020 \\
\textbf{CoNN}                           & \textbf{architecture} & \textbf{2} & \textbf{0.269} & \textbf{3558} & \textbf{0.728} & \textbf{122} & \textbf{AME2020} \\
\bottomrule
\end{tabular}
\begin{tablenotes}[flushleft]
\footnotesize
\item[a] Plain MLP with ${\sim}7.4\times 10^4$ parameters, matching the CoNN parameter count.
\end{tablenotes}
\end{threeparttable}
\end{table*}

\subsection{Training protocol}\label{sec:training}

All branches are optimized with full-batch gradients to minimize the mean-squared error on the experimental binding energies,
$\mathcal{L}(\theta)=n^{-1}\sum_{i}\left(B_{\mathrm{pred}}^{(i)}-B_{\mathrm{exp}}^{(i)}\right)^2$,
where $n$ is the number of training nuclei.
A central challenge in training an additive architecture is that the loss fixes the sum of the branch outputs rather than a unique allocation among them\,\cite{1991Jacobs+Adaptive,2020Piratla+Efficient}: the smooth branch can absorb local fluctuations, while correction branches can fit part of the global trend.
We therefore use a two-phase alternating protocol to encourage a division of labor between the smooth trend and local corrections.
Such stage-wise approaches can simplify the optimization of modular architectures by introducing different parameter groups sequentially\,\cite{2015Wright+Coordinate,2019Zeng+Global}.

In the warmup phase, the smooth macroscopic network is trained alone for 1500 full-batch epochs on the standardized target
$y_{\mathrm{std}}=(B_{\mathrm{exp}}-\mu_y)/\sigma_y$
with Adam\,\cite{2015Kingma+Adam} (learning rate $\alpha_{\mathrm{Macro}}=2\times 10^{-4}$).
This step initializes the global trend before the correction branches are introduced.
In the cooperative phase, the macroscopic and correction parameters are then updated for 400 alternating rounds:
\begin{enumerate}
    \item \emph{Macroscopic step}: the correction branches are frozen, and the macroscopic branch is trained for 50 full-batch gradient steps on $y_{\mathrm{Macro}} = (B_{\mathrm{exp}} - E_{\mathrm{emb}}^{*} - E_{\mathrm{grid}}^{*} - E_{\mathrm{pair}}^{*} - \mu_y)/\sigma_y$ with Adam ($\alpha_{\mathrm{Macro}}=2\times 10^{-4}$). Asterisks denote branch outputs held fixed during the current step.
    \item \emph{Correction step}: the macroscopic branch is frozen, and the three correction branches are trained for 100 full-batch gradient steps on the residual $y_{\mathrm{Micro}} = B_{\mathrm{exp}} - E_{\mathrm{Macro}}^{*}$ with AdamW\,\cite{2019Loshchilov+Decoupled} ($\alpha_{\mathrm{Micro}}=2\times 10^{-3}$, weight decay $10^{-3}$).
\end{enumerate}
The optimizer choice follows the regularization applied to each parameter group.
Adam is used for the smooth macroscopic network without weight decay.
AdamW is used for the discrete embeddings, regional grid, and parity-aware network so that weight decay acts as decoupled shrinkage on the local correction parameters.
The correction branches also use a larger learning rate, with $\alpha_{\mathrm{Micro}}/\alpha_{\mathrm{Macro}}=10$, so that they can adapt more rapidly to residual structure after the macroscopic update.
During the cooperative phase, both optimizers follow cosine-annealing schedules\,\cite{2017Loshchilov+SGDR}, and the best model state, selected by its performance on the validation set, is retained for final evaluation.

To reduce sensitivity to random initialization, we train an ensemble of $N_{\mathrm{ens}}=5$ models with different seeds\,\cite{2017Lakshminarayanan+Simple}.
The final prediction is the ensemble mean,
$\bar{B}_{\mathrm{pred}} = N_{\mathrm{ens}}^{-1} \sum_{k} B_{\mathrm{pred}}^{(k)}$.
We also report the ensemble standard deviation as a measure of inter-member disagreement\,\cite{2021Hullermeier+Aleatoric,2021Izmailov+What,2021Rahaman+Uncertainty}.

\section{Results and Discussion}\label{sec:results}

We evaluate CoNN in three stages: predictive accuracy and branch ablations (Sec.~\ref{sec:accuracy}); correction-module diagnostics and embedding shell-kink stability (Sec.~\ref{sec:structures}); and derived quantities and neutron-rich extrapolation (Sec.~\ref{sec:derived}).
Published direct-prediction results provide performance context, whereas the parameter-matched MLP and branch ablations provide controlled tests of the architecture.
Unless otherwise stated, CoNN and matched-MLP metrics are computed from predictions averaged over five independently trained members; literature values are reproduced as reported in the cited studies.

\subsection{Predictive performance and ablation analysis}\label{sec:accuracy}

\begin{figure*}[!t]
\centering
\includegraphics[width=\textwidth]{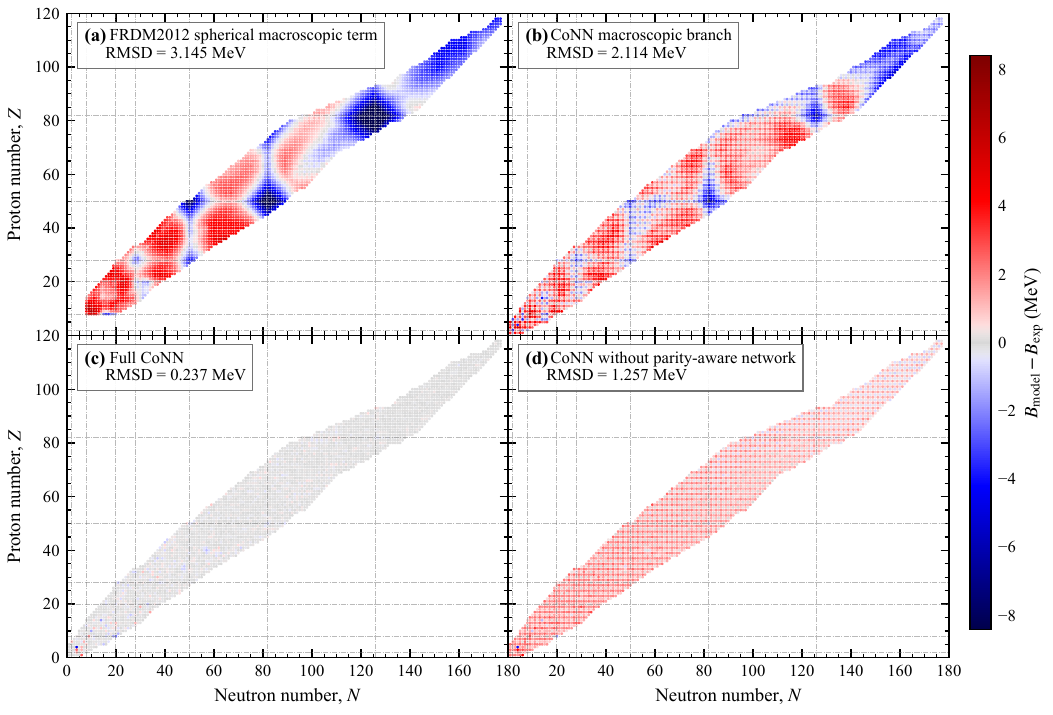}
\caption{Residual maps for branch ablations on the AME2016-overlap set ($n=3436$). Colors show each model or branch output minus $B_{\mathrm{exp}}$ (MeV) on the $(N,Z)$ chart. Panels show (a) the FRDM2012 spherical macroscopic term, (b) the CoNN macroscopic branch alone, (c) the full CoNN model, and (d) CoNN without the parity-aware network. The 122-nucleus extrapolation set is excluded.}
\label{fig:fourpanel}
\end{figure*}

Table~\ref{tab:comparison} restricts the comparison to baseline-free direct-prediction ML models and identifies where explicit nuclear physics priors enter each model.
Among models with no explicit physics prior beyond $(Z,N)$, ANN2, KAN-2, and the parameter-matched MLP give RMSDs of 1.180, 0.870, and 0.836\,MeV, respectively.
Feature-augmented direct predictors reach lower RMSDs (ANN7: 0.200\,MeV; KAN-11: 0.260\,MeV) by adding explicit nuclear descriptors\,\cite{2024Zeng+Nuclear,2025Liu+Kolmogorovarnold}.
CoNN uses the same two explicit input features as the plain models and reaches 0.269\,MeV on all 3558 nuclei in AME2020.
This value is close to KAN-11 and substantially below the other two-feature models, indicating that architecture-level physics priors recover part of the accuracy gain associated with engineered descriptors.

As a parameter-count control, we trained a plain MLP with a matched parameter budget.
This control used the same $(Z,N)$ inputs, data split, ensemble size, and total number of gradient steps as CoNN.
The parameter-matched MLP gives ensemble RMSDs of 0.836\,MeV on all 3558 AME2020 nuclei and 1.232\,MeV on the 122 extrapolation nuclei (Table~\ref{tab:comparison}).
The corresponding CoNN values are 0.269 and 0.728\,MeV, respectively, showing that matching the parameter count is not sufficient to reproduce the CoNN accuracy.

The residual maps in Fig.~\ref{fig:fourpanel} show how the branch structure changes the spatial pattern of mass residuals.
Panel (a) shows the FRDM2012 spherical macroscopic residual as a qualitative reference for global bulk structure\,\cite{2016Moller+Nuclear}.
The CoNN macroscopic branch alone gives an RMSD of 2.114\,MeV [panel (b)] and captures broad arc-like residual bands similar to those in panel (a).
After adding the discrete embeddings, regional grid, and parity-aware network, the full CoNN model reaches an RMSD of 0.237\,MeV on this AME2016-overlap set [panel (c)] and removes much of the visible systematic residual structure.
Removing the parity-aware network restores a checkerboard pattern [panel (d), RMSD\,=\,1.257\,MeV], consistent with unresolved odd--even staggering.
This ablation shows that the parity-aware branch accounts for a large and spatially widespread correction in the present architecture.

We next separate random held-out interpolation from temporal near-boundary extrapolation.
CoNN reaches 0.419\,MeV on the 20\% held-out interpolation subset ($n=688$) and 0.728\,MeV on the 122-nucleus extrapolation set.
These values are ensemble-mean predictions from five independently initialized models.
Table~\ref{tab:ensemble_spread} reports the corresponding single-member RMSDs.

\begin{table}[!htb]
\caption{Single-member and ensemble RMSDs for the five-member CoNN ensemble. ``Members'' gives the mean and standard deviation of RMSDs across individual members; ``Range'' gives the best--worst single-member interval; ``Ensemble'' gives the RMSD of the ensemble mean. All values are in MeV.}
\begin{ruledtabular}
\begin{tabular}{lcccc}
Subset & $n$ & Members & Range & Ensemble \\
\hline
Train & 2748 & $0.281\pm0.014$ & 0.256--0.297 & 0.163 \\
Validation & 688 & $0.620\pm0.017$ & 0.595--0.649 & 0.419 \\
Extrapolation & 122 & $1.162\pm0.184$ & 0.943--1.476 & 0.728 \\
Overall & 3558 & $0.428\pm0.017$ & 0.412--0.459 & 0.269 \\
\end{tabular}
\end{ruledtabular}
\label{tab:ensemble_spread}
\end{table}

The ensemble mean improves on every individual member in each subset: the overall RMSD decreases from the best single-member value of 0.412\,MeV to 0.269\,MeV, and the extrapolation RMSD decreases from 0.943\,MeV to 0.728\,MeV.
The single-member spread is small on the training, validation, and overall sets, but grows on the extrapolation set, where the best--worst range reaches 0.533\,MeV.
This larger spread provides model-disagreement information for the extrapolation diagnostics below.

Taken together, the comparison, parameter-count control, and branch ablation tests show that the CoNN architecture adds value beyond model size and raw $(Z,N)$ input.
CoNN approaches feature-augmented direct predictors in overall RMSD.
The branch ablation also shows that the correction branches play their intended roles in part.
This motivates the branch-output diagnostics in Sec.~\ref{sec:structures}, where we examine which physical structures are reflected in the learned correction components.
At the same time, the extrapolation results expose a clear limitation.
CoNN accuracy deteriorates substantially on the extrapolation set relative to the training and validation subsets.
Part of this degradation follows from the regional-grid branch, which is designed primarily to interpolate nonseparable local corrections within the measured chart.
In data-sparse boundary regions, grid values without nearby training support are weakly constrained and can contribute little learned correction.
The prediction then falls back toward the smooth macroscopic branch, the separable $Z$- and $N$-indexed embeddings, and the parity-aware correction.
This fallback preserves the imposed smooth, shell-indexed, and parity-aware structures, but loses much of the learned two-dimensional regional correction.
The extrapolation error therefore reflects a predictable architectural fallback.
To examine how this architectural fallback shapes predictions farther from measured nuclei, Sec.~\ref{sec:derived} compares the CoNN neutron-drip-line prediction with those of physics-based mass models.

\subsection{Diagnostics of microscopic correction modules}\label{sec:structures}

The correction branches are intended to represent microscopic residual structure left by the smooth macroscopic branch.
We next analyze the outputs of the three microscopic correction branches.
We consider the parity-aware network first, followed by the discrete embeddings and the regional grid.

\begin{figure}[!ht]
\centering
\includegraphics[width=\columnwidth]{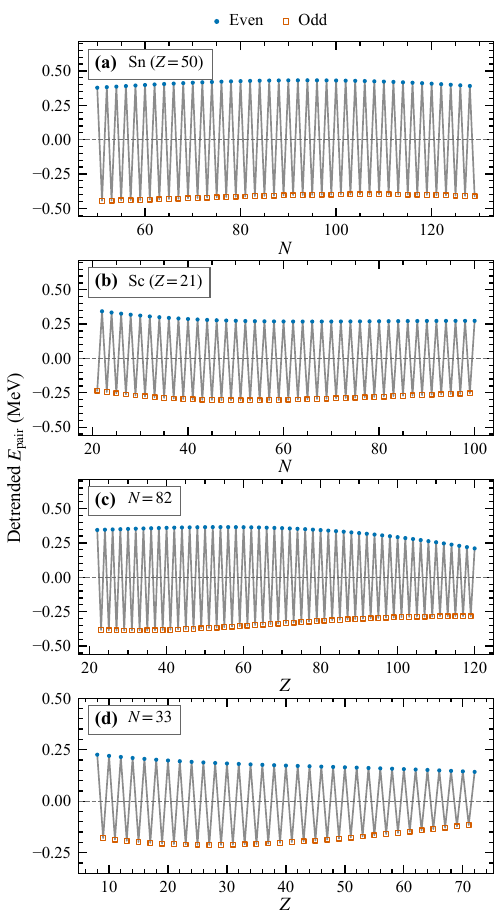}
\caption{Learned parity-aware output $E_{\mathrm{pair}}$ after removing a linear trend along each representative chain: (a) Sn and (b) Sc isotopic chains, and the (c) $N=82$ and (d) $N=33$ isotonic chains. Within each panel, filled circles and open squares denote even and odd values, respectively. The alternating pattern narrows toward heavier nuclei, consistent with the empirical decrease of odd--even staggering with mass number.}
\label{fig:pairing_sawtooth}
\end{figure}

Figure~\ref{fig:pairing_sawtooth} shows the learned parity-aware output $E_{\mathrm{pair}}$ along representative isotopic chains [panels (a) and (b), varying $N$ at fixed $Z$] and isotonic chains [panels (c) and (d), varying $Z$ at fixed $N$], after removing a smooth linear trend from each chain to isolate the oscillatory component.
In all four panels, this output displays the expected alternating sawtooth pattern, with even and odd nucleon numbers distinguished by both marker shape and color.
The staggering amplitude narrows progressively toward heavier nuclei along each chain, with the trend particularly clear in the isotonic chains in panels (c) and (d).
Using the consecutive-triplet RMS statistic from the branch-separation diagnostics, the ensemble-mean parity-aware output gives local odd--even components of 0.85\,MeV along isotopic chains and 1.06\,MeV along isotonic chains.
These values are computed on the AME2016-overlap nuclei.
The narrowing trend is consistent with empirical odd--even mass systematics, in which pairing effects generally weaken with increasing mass number\,\cite{1969Bohr+Nuclear}.

However, it should be noted that not all parity-dependent structure has been isolated in this branch.
The branch forms and alternating training protocol encourage such a division of labor, but the training targets are total binding energies and the loss constrains only the summed prediction.
Without branch-level labels or an explicit constraint that enforces a one-to-one allocation of residual structures, flexible learned components can share the same pattern\,\cite{2019Locatello+Challenging}.
Accordingly, while the parity-aware branch captures clear odd--even staggering, parity-dependent variation can also be carried by other modules, such as the discrete embeddings analyzed below.

\begin{figure}[!t]
\centering
\includegraphics[width=\columnwidth]{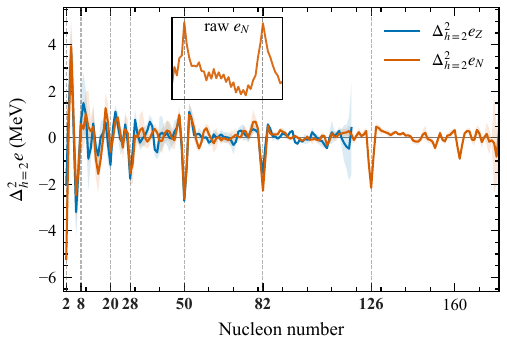}
\caption{Second differences of the learned discrete proton and neutron embeddings. Curves show $\Delta_{h=2}^2 e_Z$ and $\Delta_{h=2}^2 e_N$ as functions of nucleon number, with shaded bands denoting the standard deviation across the five ensemble members. Dashed vertical lines mark the canonical magic numbers. Negative excursions near shell closures correspond to local extra binding. The inset shows the ensemble-mean raw neutron embedding $e_N$ in the aligned window $N=45$--90.}
\label{fig:embedding_second_diff}
\end{figure}

\begin{figure*}[!t]
\centering
\includegraphics[width=\textwidth]{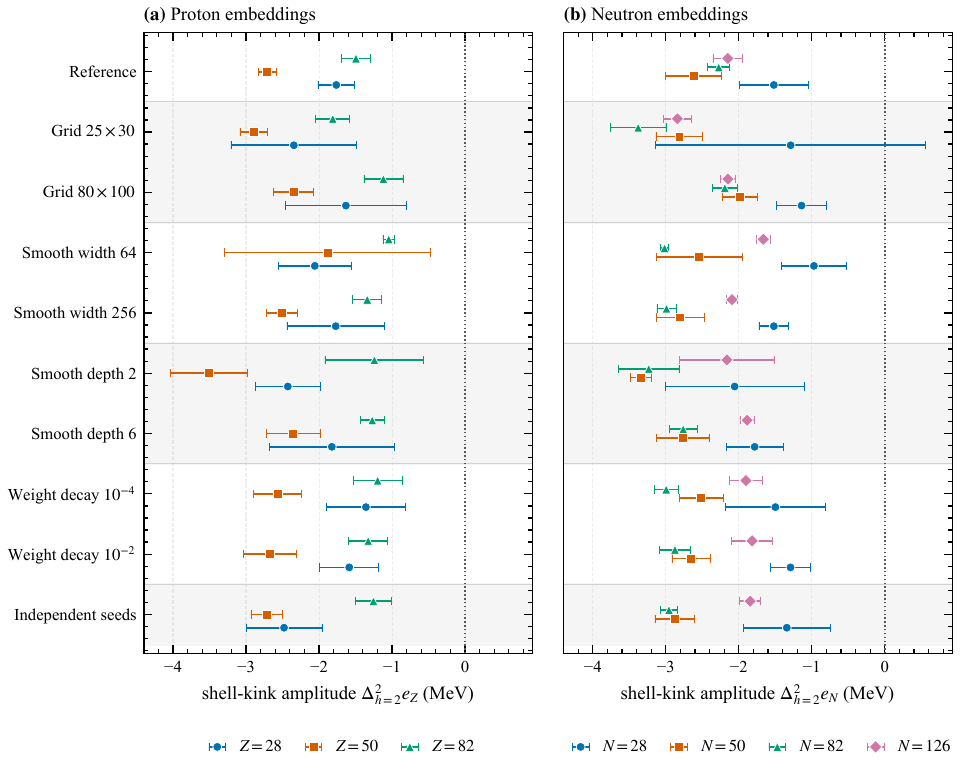}
\caption{Robustness of shell-kink features in the discrete embeddings under controlled model perturbations. The reference configuration uses a $50\times60$ regional grid, smooth-branch width 128 and depth 4, and correction-branch weight decay $10^{-3}$. The other rows vary one setting at a time relative to this reference. All configurations use cooperative training and scalar embeddings. Points show means over five ensemble members, and horizontal error bars show the standard deviations across those members within each configuration. The amplitudes are same-parity second differences $\Delta_{h=2}^2 e_Z$ and $\Delta_{h=2}^2 e_N$ at the major magic numbers. Negative values indicate local extra binding relative to the same-parity neighbors.}
\label{fig:kink_amplitudes}
\end{figure*}

The discrete embeddings were introduced as $Z$- and $N$-indexed local corrections motivated by shell closures.
We therefore ask whether the learned embedding curves contain local kink structures near magic numbers.
The inset of Fig.~\ref{fig:embedding_second_diff} shows the ensemble-mean raw neutron embedding $e_N$ over $N=45$--90.
The curve contains visible odd--even modulation, showing that some parity-dependent pattern leakage is present in the embeddings.
For the shell diagnostic, we focus on local curvature rather than absolute baselines, because constant and linear terms can be transferred between the embeddings and the smooth branch without changing their summed prediction.
Specifically, we use the step-two second difference
\begin{equation}
\Delta_h^2 e_Z[Z]=e_Z[Z+h]-2e_Z[Z]+e_Z[Z-h],
\end{equation}
with the analogous definition for $e_N$.
This quantity is unchanged by constant and linear shifts of the embedding curves.
Using $h=2$ compares embedding entries with the same nucleon-number parity and therefore reduces direct sensitivity to the odd--even modulation visible in the raw embeddings.
Figure~\ref{fig:embedding_second_diff} shows the resulting second differences of the learned proton and neutron embeddings.
Although no magic-number labels enter the training objective, negative excursions appear near the major shell closures $Z = 28, 50, 82$ and $N = 28, 50, 82, 126$.
These excursions show that the embeddings respond strongly to shell-related local structure, as intended by their $Z$- and $N$-indexed design.
They are consistent with local extra binding from shell closures associated with large gaps in the single-particle spectrum\,\cite{2024Buskirk+Nucleonic}.
For $Z,N<20$, the curves instead show high-frequency point-to-point variations.
We interpret these light-region variations as local residual structure absorbed by the embeddings, rather than as shell-closure signatures.

To test the stability of the embedding shell-kink signatures, we compared ten five-member cooperative ensembles with scalar embeddings.
The controlled variants change the regional-grid resolution, smooth-branch width or depth, correction-branch weight decay, or initialization seeds relative to the reference model.
Their validation RMSDs span 0.419--0.523\,MeV.
At each of the seven major shell closures, the configuration mean of the second difference remains negative in all ten ensembles (Fig.~\ref{fig:kink_amplitudes}).
At the member level, 348 of the 350 values are negative. The two positive values occur at $N=28$ for one member of the $25\times30$-grid ensemble and at $Z=50$ for one member of the width-64 ensemble.
Although the configuration-mean second differences remain negative, their magnitudes vary across configurations. Across the ten configuration means, $\Delta_{h=2}^2e_Z(50)$ ranges from $-3.51$ to $-1.88$\,MeV, and $\Delta_{h=2}^2e_N(82)$ ranges from $-3.38$ to $-2.19$\,MeV.
The reproducible feature is therefore the location and ensemble-level sign of the major shell kinks, whereas their precise amplitudes depend on model configuration and initialization.

\begin{figure}[!tb]
\centering
\includegraphics[width=\columnwidth]{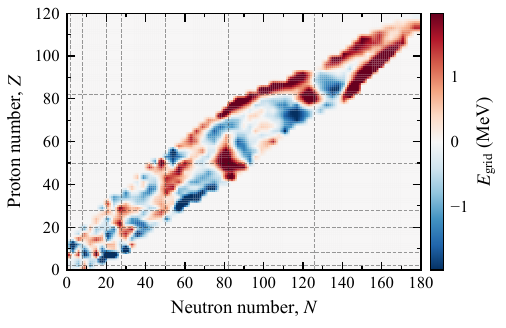}
\caption{Ensemble-mean regional-grid output $E_{\mathrm{grid}}(Z,N)$ on the $(N,Z)$ chart. Dashed lines mark magic numbers. The map contains localized patches near the doubly magic nuclei and broader structures in the rare-earth and actinide regions.}
\label{fig:grid_heatmap}
\end{figure}

The regional grid $E_{\mathrm{grid}}$ represents residual variation that depends jointly on $(Z,N)$, beyond the separable embedding contribution $e_Z+e_N$.
Figure~\ref{fig:grid_heatmap} shows the ensemble-mean regional-grid output on the $(N,Z)$ chart.
The map contains localized corrections near the doubly magic nuclei $^{132}$Sn and $^{208}$Pb, together with broader structures in the rare-earth and actinide regions.
These patterns show how the regional grid supplements the one-dimensional embeddings with nonseparable regional corrections.

Taken together, the branch outputs organize the learned correction into three recognizable patterns: odd--even staggering, index-wise shell kinks, and nonseparable regional structure.
They therefore provide a diagnostic view of how different functional forms represent the mass surface in a baseline-free fit.

\subsection{Derived quantities and extrapolation diagnostics}\label{sec:derived}

Separation energies and decay $Q$ values probe local variations of the predicted mass surface because they are formed from binding-energy differences between neighboring nuclei.
Smooth common offsets can cancel, whereas rapidly varying residuals enter directly, so a low binding-energy RMSD does not by itself ensure accurate derived observables\,\cite{2014Sobiczewski+Predictive,2025Martinet+Impact}.
Table~\ref{tab:derived} reports CoNN RMSDs for six derived quantities on the AME2016-overlap set.

\begin{table}[!htb]
\caption{RMSDs for derived quantities on the AME2016-overlap set ($n$ varies by availability of neighboring nuclei).}
\begin{ruledtabular}
\begin{tabular}{lcc}
Quantity & RMSD (MeV) & Nuclei \\
\hline
$S_n$ & 0.317 & 3317 \\
$S_{2n}$ & 0.316 & 3199 \\
$S_p$ & 0.331 & 3257 \\
$S_{2p}$ & 0.324 & 3081 \\
$Q_\alpha$ & 0.294 & 3297 \\
$Q_{\beta^-}$ & 0.360 & 3141 \\
\end{tabular}
\end{ruledtabular}
\label{tab:derived}
\end{table}

\begin{figure}[!htb]
\centering
\includegraphics[width=\columnwidth]{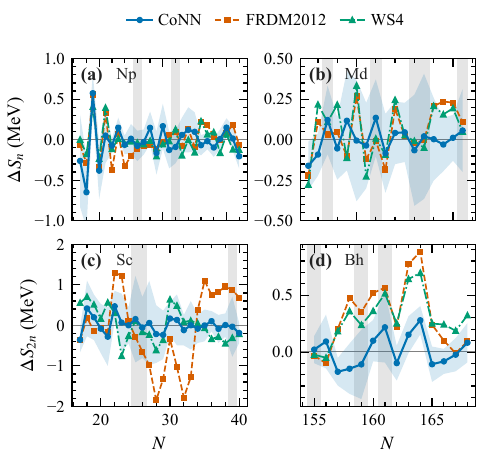}
\caption{Separation-energy residuals along selected isotopic chains: $\Delta S_n$ for Np and Md (top), $\Delta S_{2n}$ for Sc and Bh (bottom). The CoNN (blue circles) is compared with FRDM2012 (orange squares, dashed) and WS4 (green triangles, dash-dotted); the shaded blue band shows the CoNN ensemble spread. Vertical gray bands indicate nuclei not included in the training set.}
\label{fig:sep_residual}
\end{figure}

Across the six derived quantities, the RMSDs range from 0.294 to 0.360\,MeV, compared with 0.237\,MeV for binding energies on the same set.
Figure~\ref{fig:sep_residual} compares separation-energy residuals, $\Delta S=S_{\mathrm{pred}}-S_{\mathrm{exp}}$, from CoNN, FRDM2012, and WS4 along four isotopic chains.
Along the Np and Md chains, the $S_n$ residuals of all three models fluctuate around zero and remain within a few tenths of an MeV for most nuclei.
For $S_{2n}$, the contrast is clearer along Sc and Bh, where the CoNN residuals show smaller excursions than those of FRDM2012 and WS4 over the displayed nuclei.

\begin{figure}[!t]
\centering
\includegraphics[width=\columnwidth]{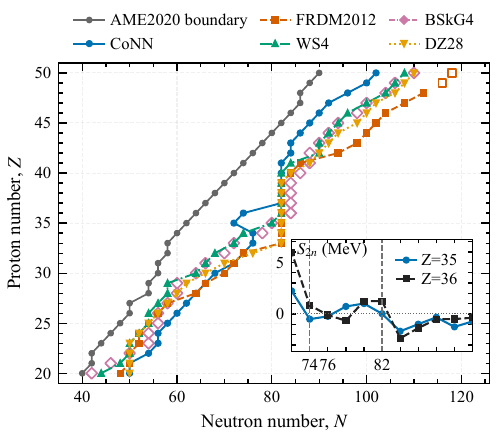}
\caption{Neutron drip-line comparison inferred from the first $S_{2n}=0$ crossing beyond the AME2020 boundary for even-$N$ isotopic chains with $20 \leq Z \leq 50$. The CoNN result is shown together with FRDM2012, WS4, BSkG4, and DZ28. The inset shows the nonmonotonic $S_{2n}$ behavior for $Z=35$ and 36 near $N=82$, which produces the local inward bend in the CoNN drip-line curve.}
\label{fig:b1_dripline}
\end{figure}

We next used $S_{2n}$ to diagnose neutron-rich extrapolation beyond the AME2020 boundary.
For even-$N$ isotopic chains with $20 \leq Z \leq 50$, the neutron drip line was defined by the first zero crossing of $S_{2n}$ beyond the largest AME2020 neutron number for that $Z$.
The CoNN drip line generally lies within the range spanned by FRDM2012\,\cite{2016Moller+Nuclear}, WS4\,\cite{2014Wang+Surface}, BSkG4\,\cite{2025Grams+Skyrme}, and DZ28\,\cite{1995Duflo+Microscopic}, but it contains a local inward bend near $Z=34$--35 (Fig.~\ref{fig:b1_dripline}).
For $Z=35$, where the AME2020 boundary is at $N=62$, the predicted $S_{2n}$ changes from 2.23\,MeV at $N=72$ to $-0.51$\,MeV at $N=74$, recovers to 0.02\,MeV at $N=82$, and becomes negative again at $N=84$.
This nonmonotonic ``dip--recover--fall'' pattern sets an early first crossing and produces the local bend in the drip-line curve.

\begin{table}[!htb]
\caption{Branch-level finite-difference contributions to $S_{2n}$ for $Z=35$ near the local bend in the CoNN neutron drip line. Here $\Delta E_M$, $\Delta e_N$, $\Delta E_g$, and $\Delta E_p\equiv\Delta E_{\mathrm{pair}}$ denote the macroscopic, neutron-embedding, regional-grid, and parity-aware contributions, respectively. All values are in MeV.}
\setlength{\tabcolsep}{4pt}
\begin{ruledtabular}
\begin{tabular}{cccccc}
$N$ & $S_{2n}$ & $\Delta E_M$ & $\Delta e_N$ & $\Delta E_g$ & $\Delta E_p$ \\
\hline
72 & 2.23 & 1.44 & $-0.16$ & 0.94 & 0.01 \\
74 & $-0.51$ & $-0.40$ & $-0.12$ & 0.00 & 0.01 \\
80 & 1.01 & 0.13 & 0.86 & 0.00 & 0.02 \\
82 & 0.02 & $-1.23$ & 1.24 & 0.00 & 0.01 \\
84 & $-1.68$ & $-0.65$ & $-1.04$ & 0.00 & 0.01 \\
\end{tabular}
\end{ruledtabular}
\label{tab:b1_decomposition}
\end{table}

The branch-level finite differences identify how this feature is generated by the current additive form.
For fixed $Z$, the proton embedding cancels in $S_{2n}(Z,N)=B(Z,N)-B(Z,N-2)$, giving
\begin{equation}
S_{2n}=\Delta E_{\mathrm{Macro}}+\Delta e_N+\Delta E_{\mathrm{grid}}+\Delta E_{\mathrm{pair}},
\end{equation}
where $\Delta f=f(Z,N)-f(Z,N-2)$.
At fixed $Z=35$, the nuclei at $N$ and $N-2$ have the same parity indicators, so $\Delta E_{\mathrm{pair}}$ contains no odd--even switch. As shown in Table~\ref{tab:b1_decomposition}, $\Delta E_{\mathrm{pair}}$ is only $0.01$--$0.02$\,MeV for these nuclei.
The initial drop from $N=72$ to 74 results from the decrease in $\Delta E_M$ and the abrupt disappearance of $\Delta E_g$.
With $\Delta E_g=0$ beyond $N=74$, the subsequent structure reflects the interplay between $\Delta E_M$ and $\Delta e_N$. Specifically, $\Delta e_N$ drops sharply from $1.24$\,MeV at $N=82$ to $-1.04$\,MeV at $N=84$, providing the dominant shell-related contribution to the fall across the $N=82$ closure.
Thus, the local bend results from the decrease in $\Delta E_M$ and the abrupt loss of regional-grid support near $N=74$, followed by the shell-related change in $\Delta e_N$ across $N=82$.

\section{Summary and Outlook}\label{sec:conclusion}

We have presented CoNN, a physics-structured modular neural network for predicting nuclear binding energies with $(Z,N)$ as its only external inputs.
Its binding-energy prediction is the sum of four form-constrained branches: a smooth macroscopic network, discrete embeddings, a two-dimensional regional grid, and a parity-aware network.
This design incorporates selected nuclear-structure priors through the branch architectures rather than through additional input descriptors.

Across all 3558 AME2020 nuclei, CoNN obtains an RMSD of 0.269\,MeV, compared with 0.836\,MeV for the parameter-matched MLP.
Because the control uses the same inputs, data split, ensemble size, training budget, and parameter count, matching model size is insufficient to reproduce CoNN's accuracy.
The RMSD nevertheless increases from 0.419\,MeV on the held-out interpolation subset to 0.728\,MeV on the 122-nucleus extrapolation set.
Branch diagnostics show odd--even staggering in the parity-aware output, embedding shell-kink signs that persist across controlled model variants, and nonseparable regional corrections.
On the AME2016-overlap set, separation energies and decay $Q$-values have RMSDs of 0.29--0.36\,MeV.
The CoNN neutron-drip-line prediction generally lies within the range of four physics-based reference models but develops a local inward bend near $Z=34$--35.

For baseline-free mass prediction, these results support architectural constraints as a complementary way to encode selected physics priors.
CoNN constrains the available functional forms, while the detailed $(Z,N)$-dependent patterns within those forms are learned from mass data rather than supplied as additional descriptors.
The branch outputs are therefore interpretable at the level of stable features rather than as a unique decomposition of the binding energy into physical terms.
The controlled model variants establish such stability specifically for the ensemble-level signs of embedding shell kinks at major magic numbers.
Odd--even modulation in both the parity-aware output and the raw embeddings shows that the allocation of learned patterns among branches remains partial.

The analyses identify three limits of the present architecture.
First, the embedding lookup tables restrict the implemented model to $Z \leq 120$ and $N \leq 180$, although this domain contains all AME2020 nuclei.
Second, within that domain, the regional grid uses local bilinear interpolation and does not learn an extrapolative continuation across regions without training support.
Grid nodes without nearby training nuclei remain weakly constrained, so $E_{\mathrm{grid}}$ can change abruptly near data boundaries and contribute little beyond them.
Predictions then increasingly rely on the macroscopic, embedding, and parity-aware branches.
This fallback contributes to the observed extrapolation degradation and to the local bend in the neutron-drip-line prediction.
Third, variation among ensemble members measures model disagreement but does not provide calibrated posterior uncertainty.

Future work should address the two extrapolation limits separately.
Continuous $Z$- and $N$-dependent corrections could remove the embedding lookup boundary, provided that they retain sharp shell-related structure.
For the regional grid, a possible extension is to weight its output according to nearby training-data coverage, so that its contribution decreases smoothly toward zero in sparsely sampled regions.
Predictive uncertainty also requires calibration beyond ensemble disagreement, for example through Bayesian methods\,\cite{2018Neufcourt+Bayesian,2020Neufcourt+Quantified,2024Saito+Uncertainty,1992MacKay+Practical}.
Extending the modular design to charge radii and $\beta$-decay observables would test which architecture-level priors remain useful beyond nuclear masses.